**Title:** International trade of flowers. Tendencies and policies

**Author:** Vítor João Pereira Domingues Martinho

Unidade de I&D do Instituto Politécnico de Viseu

Av. Cor. José Maria Vale de Andrade

Campus Politécnico

3504 - 510 Viseu

PORTUGAL

e-mail: vdmartinho@esav.ipv.pt

# International trade of flowers. Tendencies and policies

**Abstract**

There are few papers about the international trade of flowers, so it is believed that this paper, with this topic, could be an important contribution to the international scientific community. It is intended to analyze if the international trade flowers tendencies and policies are adapted to the actual world global context. For that it was used data about the import and export of flowers, in different forms, between Portugal and the world. This is an approach to understand the international trade flowers tendencies and policies. To better understand the data analyzed it is made several estimations based in the absolute convergence theory and an analyze of the data volatility. As main conclusions, there is a tendency to the countries trade the flowers between the neighbors and is needed a more coherent policy for the international trade of flowers.

# 1. Introduction

The climate in is an important resource for agrarian production, namely for the flowers productions, because, the territory of some countries, as the Mediterranean countries, has sun in the majority of the months, during the year.

Nevertheless, these productions are not totally explored in some countries landscape, because a lot of reasons.

The most important reason, in the European Union countries, is because the common agricultural policy. This European policy is totally oriented to the agricultural sector of the north countries and few oriented to the south and Mediterranean. In this way the Mediterranean farmers are induced to produce the agricultural products typical of the European north countries and not their products, like the flowers and others products from the horticulture and fruit production. The farmers are induced, because they receive more supports if they choose some productions in detriment of others. As, the agricultural productions have a lot of risks and uncertainties, the farmers opt for productions with at least one guarantee that is the public support, in form of subsidies.

In another way, it is needed some adjustments in the international and European policies for the flowers productions and commercialization and, of course, for the international trade of these productions. Because, it is needed to demonstrate to the farmers that despite the flowers productions can have few national and European supports, this sector well organized could be a profitable sector, with benefits to the rural populations and the countries.

The landscape of some countries origin small farms, but there are some works that demonstrate that is needed small area for be profitable the flowers production.

So, in this work, it is tried to analyze the data about the international trade of flowers with Portugal. This, because the international trade is the first step to become profitable the production of a sector with tradable products, like said the Keynesian theorist. From here it is analyzed, with the convergence theory and some test of stationary, the stability of the data.

This analyze is an approach to conclude about the evolution of the international flowers sector, the perspective for the future and some adjustments of policy need for become more profitable this sector with great possibility of grow in some countries. The data used are from 2006 to 2010 and were obtained from the INE (Statistics Portugal), gently given by the AICEP (Trade & Investment Agency).

The works available about the flower sector, analyze, this question, in different perspective. For example, Reinten et al. (2011) analyzed the cut flowers activity in the southern African and it potential for the international trade. The influence of the United Kingdom retailers in the international trade of flowers was analyzed by (Hughes, 2000). The relationship between the international trade of flowers and some plant pest dissemination was studied by (Areal et al., 2007). The labor conditions in the cut flowers activities were considered by Riisgaard (2008). The African flowers growers potential in the European markets were researched by Cunden and Van Heck (2004). In another way, Vringer and Blok (2000) analyzed the environment implications of the decorative cut flowers.

**2. Data analysis**

In the tables 1 and 2, with absolute values, it is presented the countries with more than ninety per cent of import and export, together, of flowers, in different forms, with Portugal.

The majority of import comes from the Europe, namely from Netherlands, Spain, France, Italy and Belgium (table 1).

**Table 1. Flowers, in different forms, import values (euros)**

| | Year | South Africa | Angola | Cape Verde | Brazil | USA | Swiss | China | Israel | Thailand | India | Germany | Belgium | Denmark | Spain | France | Netherlands | Italy | UK |
|---|---|---|---|---|---|---|---|---|---|---|---|---|---|---|---|---|---|---|---|
| Bulbs, tubers, roots, vegetation or flowers, etc., roots | 2006 | 2.157 | NA | NA | 1274 | 372 | NA | 1.647 | 7.986 | 138.327 | NA | 184.655 | 323.358 | 2.081 | 1.719.302 | 76.858 | 7.314.035 | 22.233 | 109 |
| | 2007 | NA | NA | NA | NA | NA | NA | 18.029 | 14.793 | NA | 218.944 | 420.957 | 13.790 | 1.414.294 | 24.020 | 8.108.684 | 83.434 | NA |
| | 2008 | NA | NA | NA | NA | NA | NA | 18.437 | 5.881 | NA | 172.713 | 263.807 | 5.610 | 979.793 | 121.148 | 7.380.338 | 208.148 | 5.262 |
| | 2009 | 2.044 | NA | NA | NA | NA | NA | 18.118 | 15.538 | 3.095 | 45.533 | 20.095 | 5.599 | 311.182 | 44.927 | 18.140.983 | 1.762 | NA |
| | 2010 | 799 | NA | NA | NA | NA | NA | 7.329 | 26.048 | NA | 158.095 | 44.527 | NA | 568.419 | 32.663 | 8.265.723 | 212 | NA |
| Other live plants (including roots), cuttings and slips, mushroom spawn | 2006 | 31.472 | NA | NA | 8.702 | 210.723 | NA | 33.211 | 46.549 | 2.643 | NA | 812.604 | 1.935.345 | 155.717 | 19.648.803 | 1.275.583 | 23.872.143 | 4.873.415 | 235.806 |
| | 2007 | 17.763 | NA | NA | 15.291 | 6.626 | 340 | 123.611 | 23.472 | 6.706 | NA | 958.877 | 1.200.847 | 86.945 | 20.719.930 | 1.557.055 | 24.984.372 | 4.956.950 | 296.460 |
| | 2008 | 33.954 | NA | NA | 2.232 | 38.996 | NA | 84.095 | 24.151 | 4.504 | 631 | 879.314 | 1.577.302 | 121.919 | 33.488.342 | 2.320.759 | 28.173.080 | 4.876.879 | 163.012 |
| | 2009 | 45.153 | NA | NA | 2.401 | 98.204 | NA | 43.796 | 25.149 | NA | 386 | 925.327 | 853.432 | 52.771 | 20.032.454 | 3.174.921 | 21.341.104 | 3.691.985 | 4.073 |
| | 2010 | 40.043 | NA | NA | NA | 6.831 | NA | 42.926 | 11.871 | 7.831 | 35.339 | 947.652 | 725.835 | 16.206 | 14.275.770 | 3.008.624 | 25.605.736 | 5.801.943 | 181.842 |
| Flowers and buds of p / branches / ornamental purposes, fresh, dried, etc. | 2006 | 403.975 | NA | NA | 621.467 | 747 | NA | 4.768 | NA | 3.420 | 18.119 | 742.311 | NA | 3.184.189 | 19.959 | 18.564.755 | 23.234 | 307.593 |
| | 2007 | 411.523 | NA | NA | 600.341 | 841 | NA | 11.067 | 2.263 | 9.908 | 4.552 | 58.322 | 703.287 | NA | 4.340.428 | 233.570 | 17.597.788 | 571.456 | 47.485 |
| | 2008 | 351.874 | NA | NA | 394.230 | NA | NA | 377 | 3.461 | 9.243 | 13.745 | 63.523 | 296.022 | NA | 4.511.597 | 183.583 | 17.554.637 | 341.210 | 7.330 |
| | 2009 | 94.730 | NA | NA | 269.030 | NA | NA | 10.439 | NA | 54.422 | 11.952 | 53.344 | NA | NA | 5.881.775 | 471.971 | 11.289.615 | 18.563 | NA |
| | 2010 | 32.594 | NA | NA | 120.340 | NA | NA | 7.348 | 7.350 | 56.163 | 23.714 | 83.297 | NA | NA | 3.699.776 | 1.096.432 | 14.337.539 | 38.959 | 1309 |
| Foliage, branches and other parts of plants, s / flowers / buttons, etc. | 2006 | 115.553 | NA | NA | 4.229 | 4.290 | 2.100 | 10.791 | 1.362 | NA | 72.208 | 44.484 | 111.946 | 9.682 | 694.007 | 233.900 | 2.701.125 | 39.742 | 104.062 |
| | 2007 | 100.953 | NA | NA | 13.542 | 5.583 | NA | 38.982 | NA | 66.024 | 64.743 | 24.685 | NA | 885.546 | 38.878 | 2.786.677 | 49.211 | 17.746 |
| | 2008 | 86.019 | NA | NA | 33.422 | 2.562 | NA | 56.646 | 3.864 | 165 | 25.814 | 11.323 | 98.458 | 26 | 2.183.000 | 5.744 | 2.990.013 | 8.428 | 22.888 |
| | 2009 | 23.300 | NA | NA | 66.852 | 482 | NA | 11.543 | NA | 168 | NA | 35.927 | 231.130 | NA | 1.109.228 | 4.997 | 2.125.187 | 22.937 | NA |
| | 2010 | 21.388 | NA | NA | 50.288 | 556 | NA | 6.239 | NA | 550 | 73.237 | 92.800 | 85.399 | NA | 1.460.915 | 4.768 | 2.134.191 | 61.688 | 140 |

Portugal export flowers, also, namely, to the Europe. Netherlands, Spain, France, United Kingdom, Italy and Germany are the most important destinations of the Portuguese flowers (table 2).

**Table 2. Flowers, in different forms, export values (euros)**

| | Year | South Africa | Angola | Cape Verde | Brazil | USA | Swiss | China | Israel | Thailand | India | Germany | Belgium | Denmark | Spain | France | Netherlands | Italy | UK |
|---|---|---|---|---|---|---|---|---|---|---|---|---|---|---|---|---|---|---|---|
| Bulbs, tubers, roots, vegetation or flowers, etc., roots | 2006 | NA | NA | NA | NA | NA | NA | NA | NA | NA | NA | NA | NA | NA | 12.708 | NA | 1.079.522 | NA | NA |
| | 2007 | NA | NA | NA | NA | NA | NA | NA | NA | NA | 585 | NA | NA | NA | 67.904 | 14.871 | 1.375.967 | NA | NA |
| | 2008 | NA | NA | NA | NA | NA | NA | NA | NA | NA | 620 | NA | 39 | NA | 63.563 | NA | 930.373 | NA | NA |
| | 2009 | NA | 222 | NA | NA | NA | NA | NA | NA | NA | NA | NA | NA | NA | 3.230.676 | NA | 780.195 | NA | NA |
| | 2010 | NA | 804 | NA | NA | NA | NA | NA | NA | NA | NA | 1250 | NA | NA | NA | 10.280 | 1.090.755 | NA | NA |
| Other live plants (including roots), cuttings and slips, mushroom spawn | 2006 | NA | 2.009 | 24.638 | 3.586 | 6.014 | 152.809 | NA | NA | NA | NA | 2.520.958 | 524.812 | 95 | 5.228.290 | 5.094.387 | 7.853.049 | 339.464 | 2.523.495 |
| | 2007 | 14.629 | 20.104 | 5.367 | 3.085 | 188.855 | 78.588 | 18.110 | 1.216 | NA | NA | 2.404.217 | 566.594 | NA | 8.195.069 | 5.416.216 | 9.087.990 | 509.067 | 2.694.163 |
| | 2008 | 90 | 68.185 | 5.808 | 5.031 | 69.720 | 73.130 | NA | 1.265 | NA | NA | 1.168.196 | 556.362 | 13.743 | 6.155.503 | 5.717.727 | 9.803.493 | 787.499 | 2.527.302 |
| | 2009 | NA | 71.048 | 3.838 | NA | 77.862 | 44.287 | NA | NA | NA | NA | 1.131.320 | 1.030.193 | 91.816 | 4.954.271 | 6.197.957 | 9.432.277 | 1.400.801 | 2.219.833 |
| | 2010 | NA | 57.885 | 3.484 | NA | 161.294 | 29.904 | NA | NA | NA | NA | 456.796 | 716.690 | 102.716 | 5.046.533 | 6.361.183 | 11.731.354 | 2.948.997 | 3.061.114 |
| Flowers and buds of p / branches / ornamental purposes, fresh, dried, etc. | 2006 | NA | 17.738 | 7.169 | NA | NA | 1572 | NA | NA | NA | NA | 314.610 | 163.381 | NA | 162.320 | 175.185 | 4.724.390 | 130.188 | 5.819 |
| | 2007 | NA | 22.538 | 3.228 | NA | NA | 1.370 | NA | NA | NA | NA | 243.326 | 21.188 | NA | 289.375 | 210.492 | 4.751.432 | 13.529 | NA |
| | 2008 | NA | 10.987 | 1.260 | NA | NA | 877 | NA | NA | NA | NA | 253.031 | 277.834 | NA | 318.687 | 6.004 | 5.256.735 | NA | 27.029 |
| | 2009 | NA | 11.726 | 3.822 | NA | NA | 1.368 | NA | 840 | NA | NA | 63.184 | 331.742 | NA | 618.525 | 35.044 | 7.508.691 | NA | 153.114 |
| | 2010 | NA | 50.352 | 11.272 | NA | NA | 1.152 | NA | NA | NA | NA | NA | 123.427 | NA | 330.097 | 3.537 | 3.213.584 | NA | 94.047 |
| Foliage, branches and other parts of plants, s / flowers / buttons, etc. | 2006 | NA | 1.360 | 602 | NA | NA | NA | NA | NA | NA | NA | 504.076 | 202.770 | NA | 12.496.917 | 603.765 | 2.196.748 | 2.099.778 | 98.945 |
| | 2007 | NA | 2.631 | NA | NA | NA | 20 | NA | NA | NA | NA | 539.568 | 228.525 | NA | 15.937.919 | 58.797 | 1.981.932 | 4.505.754 | 102.114 |
| | 2008 | NA | 4.785 | NA | NA | NA | NA | NA | NA | NA | NA | 576.795 | 262.566 | NA | 8.864.672 | 38.611 | 2.467.610 | 1.549.003 | 208.783 |
| | 2009 | NA | 4.344 | 31 | NA | NA | NA | NA | NA | NA | NA | 387.664 | 289.549 | NA | 9.353.731 | 44.199 | 1.806.112 | 2.468.492 | 156.980 |
| | 2010 | NA | 5.439 | 1.696 | NA | NA | NA | NA | NA | NA | NA | 636.750 | 296.178 | NA | 16.586.061 | 30.447 | 1.294.269 | 5.235.820 | 131.597 |

From South Africa Portugal import particularly other live plants (including roots), cuttings and slips and mushroom spawn, and flowers and buds cut for branches and ornamental purposes, fresh, dried, etc. (table 3). Do not import any flowers from Angola and Cape Verde. From Brazil import in particular flowers and buds cut for branches and ornamental purposes, fresh, dried, etc. The United States of America send to Portugal namely other live plants (including roots), cuttings and slips and mushroom spawn. The Swiss few importance has in the Portuguese flowers importations. The majority of imports from China are other live plants (including roots), cuttings and slips and mushroom spawn. From Israel are bulbs, tubers, roots, vegetation or flowers, etc. and roots and other live plants (including roots), cuttings and slips and mushroom spawn. From Thailand the majority are flowers and buds cut for branches and ornamental purposes, fresh, dried, etc. From India Portugal import namely flowers and buds cut for branches and ornamental purposes, fresh, dried, etc. and foliage, branches and other parts of plants, without flowers and buttons, etc. From the European countries Portugal import in particular other live plants (including roots), cuttings and slips and mushroom spawn. The differences in the import of flowers from different countries are normal, because their availability depend from their natural and climate conditions.

**Table 3. Flowers, in different forms, import percentage relatively to the total of each country**

| | Year | South Africa | Angola | Cape Verde | Brazil | USA | Swiss | China | Israel | Thailand | India | Germany | Belgium | Denmark | Spain | France | Netherlands | Italy | UK |
|---|---|---|---|---|---|---|---|---|---|---|---|---|---|---|---|---|---|---|---|
| Bulbs, tubers, roots, vegetation or flowers, etc., roots | 2006 | NA | NA | NA | NA | NA | NA | NA | 41,20 | 47,10 | NA | 16,83 | 17,91 | 13,69 | 5,17 | 1,30 | 15,16 | 1,47 | NA |
| | 2007 | NA | NA | NA | NA | NA | NA | NA | 36,94 | 29,71 | NA | 15,33 | 11,80 | 4,40 | 2,38 | 4,60 | 13,16 | 3,83 | 2,65 |
| | 2008 | 1,24 | NA | NA | NA | NA | NA | NA | 41,87 | 22,16 | 20,05 | 4,30 | 1,82 | 9,59 | 1,14 | 1,22 | 34,29 | 0,05 | NA |
| | 2009 | 0,84 | NA | NA | NA | NA | NA | NA | 27,60 | 28,75 | NA | 12,33 | 5,20 | NA | 2,84 | 0,79 | 16,42 | 0,00 | NA |
| | 2010 | NA | NA | NA | NA | NA | NA | NA | 4,95 | NA | NA | 9,50 | 24,19 | NA | 3,79 | 2,07 | 18,35 | 0,01 | NA |
| Other live plants (including roots), cuttings and slips, mushroom spawn | 2006 | 3,35 | NA | NA | 2,43 | 50,77 | 100,00 | 71,18 | 53,63 | 21,35 | NA | 73,71 | 51,10 | 86,31 | 75,73 | 84,01 | 46,72 | 87,56 | 81,96 |
| | 2007 | 7,20 | NA | NA | 0,52 | 93,84 | NA | 59,59 | 48,39 | 22,76 | 1,57 | 78,03 | 70,55 | 95,58 | 81,36 | 88,20 | 50,22 | 89,74 | 82,13 |
| | 2008 | 27,33 | NA | NA | 0,71 | 99,51 | NA | 66,58 | 58,13 | NA | 2,50 | 87,28 | 77,26 | 90,41 | 73,29 | 85,88 | 40,34 | 98,84 | 100,00 |
| | 2009 | 42,23 | NA | NA | NA | 92,47 | NA | 75,96 | 44,71 | 8,64 | 26,71 | 73,93 | 84,82 | 100,00 | 71,36 | 72,63 | 50,86 | 98,29 | 99,21 |
| | 2010 | 85,74 | NA | NA | 12,32 | 99,47 | NA | 27,16 | 95,05 | 3,73 | 47,86 | 80,52 | 55,27 | 97,77 | 69,26 | 84,48 | 44,94 | 92,16 | 95,54 |
| Flowers and buds of p / branches / ornamental purposes, fresh, dried, etc. | 2006 | 77,61 | NA | NA | 95,42 | 6,44 | NA | 6,37 | 5,17 | 31,55 | 6,45 | 4,48 | 29,93 | NA | 15,86 | 12,60 | 32,91 | 10,09 | 13,13 |
| | 2007 | 74,57 | NA | NA | 91,71 | NA | NA | 0,27 | 6,93 | 46,70 | 34,20 | 5,64 | 13,24 | NA | 10,96 | 6,98 | 31,29 | 6,28 | 3,69 |
| | 2008 | 57,33 | NA | NA | 79,53 | NA | NA | 15,87 | NA | 77,60 | 77,44 | 5,03 | NA | NA | 21,52 | 12,77 | 21,34 | 0,50 | NA |
| | 2009 | 34,37 | NA | NA | 70,53 | NA | NA | 13,00 | 27,68 | 62,00 | 17,93 | 6,50 | NA | NA | 18,49 | 26,47 | 28,48 | 0,66 | 0,71 |
| | 2010 | 8,06 | NA | NA | 86,52 | NA | NA | 0,12 | NA | 96,23 | NA | 2,86 | 19,31 | NA | 17,43 | 13,28 | 30,10 | 2,02 | 2,02 |
| Foliage, branches and other parts of plants, s / flowers / buttons, etc. | 2006 | 19,04 | NA | NA | 2,15 | 42,78 | NA | 22,45 | NA | NA | 93,55 | 4,98 | 1,05 | NA | 3,24 | 2,10 | 5,21 | 0,87 | 4,91 |
| | 2007 | 18,23 | NA | NA | 7,77 | 6,16 | NA | 40,14 | 7,74 | 0,83 | 64,23 | 1,00 | 4,40 | 0,02 | 5,30 | 0,22 | 5,33 | 0,16 | 11,53 |
| | 2008 | 14,10 | NA | NA | 19,76 | 0,49 | NA | 17,55 | NA | 0,24 | NA | 3,39 | 20,92 | NA | 4,06 | 0,14 | 4,02 | 0,61 | NA |
| | 2009 | 22,56 | NA | NA | 29,47 | 7,53 | NA | 11,04 | NA | 0,61 | 55,36 | 7,24 | 9,98 | NA | 7,30 | 0,12 | 4,24 | 1,05 | 0,08 |
| | 2010 | 6,20 | NA | NA | 1,16 | 0,53 | NA | 72,73 | NA | 0,03 | 52,14 | 7,12 | 1,23 | 2,23 | 9,52 | 0,17 | 6,61 | 5,81 | 2,44 |

Portugal exports namely other live plants (including roots), cuttings and slips and mushroom spawn. However the Portuguese export has little importance to South Africa, Brazil, United States of America, China, Israel, India, Thailand and Denmark (table 4).

**Table 4. Flowers, in different forms, export percentage relatively to the total of each country**

| | Year | South Africa | Angola | Cape Verde | Brazil | USA | Swiss | China | Israel | Thailand | India | Germany | Belgium | Denmark | Spain | France | Netherlands | Italy | UK |
|---|---|---|---|---|---|---|---|---|---|---|---|---|---|---|---|---|---|---|---|
| Bulbs, tubers, roots, vegetation or flowers, etc., roots | 2006 | NA | NA | NA | NA | NA | NA | NA | NA | NA | 100,00 | NA | NA | NA | 0,28 | 0,26 | 8,00 | NA | NA |
| | 2007 | NA | NA | NA | NA | NA | NA | NA | NA | NA | NA | NA | 0,00 | NA | 0,41 | NA | 5,04 | NA | NA |
| | 2008 | NA | 0,25 | NA | NA | NA | NA | NA | NA | NA | NA | NA | NA | NA | 17,79 | NA | 4,00 | NA | NA |
| | 2009 | NA | 0,70 | NA | NA | NA | NA | NA | NA | NA | NA | 0,11 | NA | NA | NA | 0,16 | 6,29 | NA | NA |
| | 2010 | NA | 1,02 | NA | NA | NA | NA | NA | NA | NA | NA | 0,03 | NA | NA | 0,00 | NA | 3,80 | NA | NA |
| Other live plants (including roots), cuttings and slips, mushroom spawn | 2006 | 100,00 | 44,41 | 62,44 | 100,00 | 100,00 | 98,26 | 100,00 | 100,00 | NA | NA | 75,44 | 69,41 | NA | 33,46 | 95,02 | 52,85 | 10,12 | 96,35 |
| | 2007 | 100,00 | 81,21 | 82,17 | 100,00 | 100,00 | 98,81 | NA | 100,00 | NA | NA | 58,47 | 50,73 | 100,00 | 39,96 | 99,23 | 53,11 | 33,70 | 91,47 |
| | 2008 | NA | 81,35 | 49,90 | NA | 98,27 | 98,14 | NA | NA | NA | NA | 71,50 | 62,38 | 100,00 | 27,29 | 98,74 | 48,30 | 36,20 | 87,74 |
| | 2009 | NA | 50,56 | 21,18 | NA | 100,00 | 96,29 | NA | NA | NA | NA | 41,72 | 63,07 | 100,00 | 22,98 | 99,31 | 67,69 | 36,03 | 93,13 |
| | 2010 | NA | 59,94 | 44,58 | NA | 100,00 | 98,91 | NA | 100,00 | NA | 100,00 | 35,66 | 62,07 | 100,00 | 22,42 | 99,51 | 74,20 | 24,13 | 98,17 |
| Flowers and buds of p / branches / ornamental purposes, fresh, dried, etc. | 2006 | NA | 49,78 | 37,56 | NA | NA | 1,71 | NA | NA | NA | NA | 7,63 | 2,60 | NA | 1,18 | 3,69 | 27,63 | 0,27 | NA |
| | 2007 | NA | 13,09 | 17,83 | NA | NA | 1,19 | NA | NA | NA | NA | 12,66 | 25,33 | NA | 2,07 | 0,10 | 28,48 | NA | 0,98 |
| | 2008 | NA | 13,43 | 49,69 | NA | 1,73 | 1,86 | NA | NA | NA | NA | 3,99 | 20,09 | NA | 3,41 | 0,56 | 38,45 | NA | 6,05 |
| | 2009 | NA | 43,98 | 68,51 | NA | NA | 3,71 | NA | NA | NA | NA | 10,86 | NA | NA | 1,50 | 0,06 | 18,54 | NA | 2,86 |
| | 2010 | NA | 24,26 | 54,08 | NA | NA | 1,09 | NA | NA | NA | NA | NA | 19,22 | NA | 5,15 | NA | 13,06 | NA | 0,48 |
| Foliage, branches and other parts of plants, s / flowers / buttons, etc. | 2006 | NA | 5,81 | NA | NA | NA | 0,03 | NA | NA | NA | NA | 16,93 | 27,99 | NA | 65,08 | 1,03 | 11,52 | 89,61 | 3,65 |
| | 2007 | NA | 5,70 | NA | NA | NA | NA | NA | NA | NA | NA | 28,87 | 23,94 | NA | 57,55 | 0,67 | 13,37 | 66,30 | 7,56 |
| | 2008 | NA | 4,97 | 0,40 | NA | NA | NA | NA | NA | NA | NA | 24,50 | 17,53 | NA | 51,52 | 0,70 | 9,25 | 63,80 | 6,20 |
| | 2009 | NA | 4,75 | 10,31 | NA | NA | NA | NA | NA | NA | NA | 58,16 | 26,07 | NA | 75,52 | 0,48 | 7,47 | 63,97 | 4,00 |
| | 2010 | NA | 14,77 | 1,34 | NA | NA | NA | NA | NA | NA | NA | 64,32 | 18,71 | NA | 72,43 | 0,49 | 8,94 | 75,87 | 1,35 |

The most important country for the Portuguese imports of flowers is clearly the Netherlands and after Spain (table 5).

**Table 5. Flowers, in different forms, import percentage relatively to the total of each year**

| | Year | South Africa | Angola | Cape Verde | Brazil | USA | Swiss | China | Israel | Thailand | India | Germany | Belgium | Denmark | Spain | France | Netherlands | Italy | UK |
|---|---|---|---|---|---|---|---|---|---|---|---|---|---|---|---|---|---|---|---|
| Bulbs, tubers, roots, vegetation or flowers, etc., roots | 2006 | 0,02 | NA | NA | 0,01 | 0,00 | NA | 0,02 | 0,08 | 1,41 | NA | 1,88 | 3,30 | 0,02 | 17,52 | 0,78 | 74,55 | 0,23 | 0,00 |
| | 2007 | NA | NA | NA | NA | NA | NA | NA | 0,17 | 0,14 | 0,02 | 2,12 | 4,08 | 0,13 | 13,70 | 0,23 | 78,55 | 0,81 | NA |
| | 2008 | NA | NA | NA | NA | NA | NA | NA | 0,20 | 0,06 | NA | 1,89 | 2,88 | 0,06 | 10,69 | 1,32 | 80,55 | 2,27 | 0,06 |
| | 2009 | 0,01 | NA | NA | NA | NA | NA | NA | 0,10 | 0,08 | 0,02 | 0,24 | 0,11 | 0,03 | 1,67 | 0,24 | 97,48 | 0,01 | NA |
| | 2010 | 0,01 | NA | NA | NA | NA | NA | NA | 0,08 | 0,29 | NA | 1,74 | 0,49 | NA | 6,24 | 0,36 | 90,79 | 0,00 | NA |
| Other live plants (including roots), cuttings and slips, mushroom spawn | 2006 | 0,06 | NA | NA | 0,02 | 0,39 | NA | 0,06 | 0,09 | 0,00 | NA | 1,52 | 3,62 | 0,29 | 36,75 | 2,39 | 44,65 | 9,11 | 0,44 |
| | 2007 | 0,03 | NA | NA | 0,03 | 0,01 | 0,00 | 0,22 | 0,04 | 0,01 | NA | 1,71 | 2,14 | 0,16 | 37,01 | 2,78 | 44,63 | 8,85 | 0,53 |
| | 2008 | 0,05 | NA | NA | 0,00 | 0,05 | NA | 0,12 | 0,03 | 0,01 | 0,00 | 1,21 | 2,17 | 0,17 | 46,04 | 3,19 | 38,73 | 6,70 | 0,22 |
| | 2009 | 0,09 | NA | NA | 0,00 | 0,19 | NA | 0,09 | 0,05 | NA | 0,00 | 1,82 | 1,68 | 0,10 | 39,44 | 6,25 | 42,01 | 7,27 | 0,01 |
| | 2010 | 0,08 | NA | NA | 0,00 | 0,01 | NA | 0,08 | 0,02 | 0,02 | 0,07 | 1,85 | 1,41 | 0,03 | 27,82 | 5,86 | 49,91 | 11,31 | 0,35 |
| Flowers and buds of p / branches / ornamental purposes, fresh, dried, etc. | 2006 | 1,63 | NA | NA | 2,52 | NA | NA | 0,02 | 0,00 | NA | 0,01 | 0,07 | 3,00 | NA | 12,89 | 0,08 | 75,14 | 0,09 | 1,24 |
| | 2007 | 1,60 | NA | NA | 2,33 | 0,00 | NA | 0,04 | 0,01 | 0,04 | 0,03 | 0,23 | 2,73 | NA | 16,87 | 0,91 | 68,40 | 2,22 | 0,18 |
| | 2008 | 1,40 | NA | NA | 1,57 | NA | NA | 0,00 | 0,01 | 0,04 | 0,05 | 0,25 | 1,18 | NA | 17,94 | 0,73 | 69,79 | 1,36 | 0,03 |
| | 2009 | 0,50 | NA | NA | 1,41 | NA | NA | 0,05 | NA | 0,28 | 0,06 | 0,28 | NA | NA | 30,74 | 2,47 | 59,01 | 0,10 | NA |
| | 2010 | 0,16 | NA | NA | 0,59 | NA | NA | 0,04 | 0,04 | 0,28 | 0,12 | 0,41 | NA | NA | 18,27 | 5,42 | 70,82 | 0,19 | 0,01 |
| Foliage, branches and other parts of plants, s / flowers / buttons, etc. | 2006 | 2,75 | NA | NA | 0,10 | 0,10 | 0,05 | 0,26 | 0,03 | NA | 1,72 | 1,06 | 2,66 | 0,23 | 16,51 | 5,56 | 64,25 | 0,95 | 2,48 |
| | 2007 | 2,40 | NA | NA | 0,32 | 0,13 | NA | 0,93 | NA | NA | 1,57 | 1,54 | 0,59 | NA | 21,04 | 0,92 | 66,21 | 1,17 | 0,42 |
| | 2008 | 1,53 | NA | NA | 0,60 | 0,05 | NA | 1,01 | 0,07 | NA | 0,46 | 0,20 | 1,75 | 0,00 | 38,90 | 0,10 | 53,28 | 0,15 | 0,41 |
| | 2009 | 0,63 | NA | NA | 1,82 | 0,01 | NA | 0,31 | NA | 0,00 | NA | 0,98 | 6,30 | 0,14 | 30,22 | 0,14 | 57,89 | 0,62 | NA |
| | 2010 | 0,53 | NA | NA | 1,24 | 0,01 | NA | 0,15 | NA | 0,01 | 1,81 | 2,30 | 2,11 | NA | 36,13 | 0,12 | 52,78 | 1,53 | 0,00 |

It can say the same to the Portuguese exports. The principal destination of the Portuguese flowers is the Netherlands and after Spain. However Spain is the first destination of Portuguese foliage, branches and other parts of plants, without flowers and buttons, etc (table 6).

**Table 6. Flowers, in different forms, export percentage relatively to the total of each year**

| | Year | South Africa | Angola | Cape Verde | Brazil | USA | Swiss | China | Israel | Thailand | India | Germany | Belgium | Denmark | Spain | France | Netherlands | Italy | UK |
|---|---|---|---|---|---|---|---|---|---|---|---|---|---|---|---|---|---|---|---|
| Bulbs, tubers, roots, vegetation or flowers, etc., roots | 2006 | NA | NA | NA | NA | NA | NA | NA | NA | NA | NA | NA | NA | NA | 1,16 | NA | 98,51 | NA | NA |
| | 2007 | NA | NA | NA | NA | NA | NA | NA | NA | NA | 0,04 | NA | NA | NA | 4,57 | 1,00 | 92,65 | NA | NA |
| | 2008 | NA | NA | NA | NA | NA | NA | NA | NA | NA | 0,06 | NA | 0,00 | NA | 6,39 | NA | 93,54 | NA | NA |
| | 2009 | NA | 0,01 | NA | NA | NA | NA | NA | NA | NA | NA | NA | NA | NA | 80,54 | NA | 19,45 | NA | NA |
| | 2010 | NA | 0,07 | NA | NA | NA | NA | NA | NA | NA | NA | NA | NA | NA | NA | 0,93 | 98,26 | NA | NA |
| Other live plants (including roots), cuttings and slips, mushroom spawn | 2006 | 0,01 | 0,10 | 0,01 | 0,02 | 0,61 | 0,81 | 0,05 | NA | NA | NA | 10,11 | 2,10 | 0,00 | 20,96 | 20,42 | 31,48 | 1,36 | 10,12 |
| | 2007 | 0,05 | 0,07 | 0,02 | 0,01 | 0,64 | 0,26 | 0,06 | 0,00 | NA | NA | 8,09 | 1,91 | NA | 27,58 | 18,23 | 30,59 | 1,71 | 9,07 |
| | 2008 | 0,00 | 0,25 | 0,02 | 0,02 | 0,25 | 0,27 | NA | 0,00 | NA | NA | 4,26 | 2,03 | 0,05 | 22,46 | 20,87 | 35,78 | 2,87 | 9,22 |
| | 2009 | 0,27 | 0,01 | NA | 0,29 | 0,17 | NA | NA | NA | NA | NA | 4,22 | 3,84 | 0,34 | 18,49 | 23,13 | 35,20 | 5,23 | 8,28 |
| | 2010 | 0,19 | 0,01 | NA | 0,52 | 0,10 | NA | NA | NA | NA | NA | 1,48 | 2,33 | 0,33 | 16,40 | 20,67 | 38,12 | 9,58 | 9,95 |
| Flowers and buds of p / branches / ornamental purposes, fresh, dried, etc. | 2006 | NA | 0,31 | 0,12 | NA | 0,03 | NA | NA | NA | NA | NA | 5,47 | 2,84 | NA | 2,82 | 3,05 | 82,13 | 2,26 | 0,10 |
| | 2007 | NA | 0,40 | 0,06 | NA | 0,02 | NA | NA | NA | NA | NA | 4,34 | 0,38 | NA | 5,16 | 3,75 | 84,68 | 0,24 | NA |
| | 2008 | NA | 0,18 | 0,02 | NA | 0,01 | NA | NA | NA | NA | NA | 4,11 | 4,51 | NA | 5,18 | 0,10 | 85,37 | NA | 0,44 |
| | 2009 | NA | 0,13 | 0,04 | NA | 0,02 | 0,01 | NA | NA | NA | NA | 0,72 | 3,79 | NA | 7,07 | 0,40 | 85,81 | NA | 1,75 |
| | 2010 | NA | 1,30 | 0,29 | NA | NA | 0,03 | NA | NA | NA | NA | NA | 3,20 | NA | 8,55 | 0,09 | 83,23 | NA | 2,44 |
| Foliage, branches and other parts of plants, s / flowers / buttons, etc. | 2006 | NA | 0,01 | 0,00 | NA | NA | NA | NA | NA | NA | NA | 2,77 | 1,11 | NA | 68,62 | 3,32 | 12,06 | 11,53 | 0,54 |
| | 2007 | NA | 0,01 | NA | NA | 0,00 | NA | NA | NA | NA | NA | 2,31 | 0,98 | NA | 68,22 | 0,25 | 8,48 | 19,29 | 0,44 |
| | 2008 | NA | 0,03 | NA | NA | NA | NA | NA | NA | NA | NA | 4,12 | 1,88 | NA | 63,36 | 0,28 | 17,64 | 11,07 | 1,49 |
| | 2009 | NA | 0,03 | 0,00 | NA | NA | NA | NA | NA | NA | NA | 2,67 | 1,99 | NA | 64,30 | 0,30 | 12,42 | 16,97 | 1,08 |
| | 2010 | NA | 0,02 | 0,01 | NA | NA | NA | NA | NA | NA | NA | 2,62 | 1,22 | NA | 68,34 | 0,13 | 5,33 | 21,57 | 0,54 |

## 3. Estimations results for the neoclassical model with panel data

The estimations results presented in the following tables, were obtained with the informatics program stata12, and are obtained with different method of estimation. Are presented all values for the different methods and the statistics tests. The statistics tests allow us to find the correct method and the correct estimation values.

The model used is the traditional model of the neoclassical theory (absolute convergence) of Solow (1956) with the development to panel data of Islam (1995).

In the table 7, presented below, it is observed the importance of the fixed effects, analyzing the statistics tests and the value of the constant coefficient. In another way, it is found strong signs of convergence observing the coefficients obtained with the static and dynamic panel data econometrics methods.

**Table 7. Results from the absolute convergence model for flowers import (absolute values)**

|  | Const.[1] | Coef.[2] | F/Wald(mod.)[3] | F(Fe_OLS)[4] | Corr(u_i)[5] | F(Re_OLS)[6] | Hausman[7] | $R^2$ [8] | N.O.[9] | N.I.[10] |
|---|---|---|---|---|---|---|---|---|---|---|
| **Bulbs, tubers, roots, vegetation or flowers, etc. and roots** | | | | | | | | | | |
| FE[11] | 7.077* (2.730) | -0.645* (-2.850) | 8.140* | 1.230 | -0.890 | ------- | ------- | 0.246 | 36 | ------- |
| RE[12] | 0.521 (0.490) | -0.070 (-0.780) | 0.600 | ------- | ------- | 0.000 | 7.710* |  | 36 | ------- |
| OLS | 0.521 (0.490) | -0.070 (-0.780) | 0.600 | ------- | ------- | ------- | ------- | 0.012 | 36 | ------- |
| DPD[13] | 21.938* (3.730) | -1.922* (-3.820) | 18.020* | ------- | ------- | ------- | ------- | ------- | 17 | 5 |
| **Other live plants (including roots), cuttings and slips and mushroom spawn** | | | | | | | | | | |
| FE[11] | 15.275* (6.640) | -1.235* (-6.670) | 44.550* | 3.990* | -0.963 | ------- | ------- | 0.533 | 55 | ------- |
| RE[12] | 1.594** (1.880) | -0.133* (-1.990) | 3.970* | ------- | ------- | 1.440 | 40.760* | 0.533 | 55 | ------- |
| OLS | ------- | ------- | ------- | ------- | ------- | ------- | ------- | ------- | ------- | ------- |
| DPD[13] | 29.127* (6.130) | -2.324* (-6.160) | 71.810* | ------- | ------- | ------- | ------- | ------- | 25 | 5 |
| **Flowers and buds cut for branches and ornamental purposes, fresh, dried, etc.** | | | | | | | | | | |
| FE[11] | 10.231* (5.050) | -0.867* (-5.030) | 25.300* | 2.900* | -0.914 | ------- | ------- | 0.450 | 45 | ------- |
| RE[12] | 2.572* (2.650) | -0.217* (-2.680) | 7.170* | ------- | ------- | 0.210 | 18.250* | 0.450 | 45 | ------- |
| OLS | ------- | ------- | ------- | ------- | ------- | ------- | ------- | ------- | ------- | ------- |
| DPD[13] | 17.944* (4.180) | -1.518* (-4.200) | 39.410* | ------- | ------- | ------- | ------- | ------- | 19 | 5 |
| **Foliage, branches and other parts of plants, without flowers and buttons, etc.** | | | | | | | | | | |
| FE[11] | 6.972* (4.530) | -0.664* (-4.620) | 21.360* | 2.790* | -0.895 | ------- | ------- | 0.400 | 46 | ------- |
| RE[12] | 1.047 (1.440) | -0.110 (-1.650) | 2.720* | ------- | ------- | 0.000 | 18.960* | 0.400 | 46 | ------- |
| OLS | ------- | ------- | ------- | ------- | ------- | ------- | ------- | ------- | ------- | ------- |
| DPD[13] | 13.480* (6.260) | -1.268* (-6.350) | 47.120* | ------- | ------- | ------- | ------- | ------- | 20 | 5 |

Note: 1, Constant; 2, Coefficient; 3, Test F for fixed effects model and test Wald for random effects and dynamic panel data models; 4, Test F for fixed effects or OLS (Ho is OLS); 5, Correlation between errors and regressors in fixed effects; 6, Test F for random effects or OLS (Ho is OLS); 7, Hausman test (Ho is GLS); 8, R square; 9, Number of observations; 10, Number of instruments;, 11, Fixed effects model; 12, Random effects model; 13, Dynamic panel data model; *, Statically significant at 5%.

In the table 8, it is verified more or less the same referred for the table 7. To stress the results of the first estimations which is because a number of observations problem.

**Table 8. Results from the absolute convergence model for flowers export (absolute values)**

|  | Const.[1] | Coef.[2] | F/Wald(mod.)[3] | F(Fe_OLS)[4] | Corr(u_i)[5] | F(Re_OLS)[6] | Hausman[7] | $R^2$ [8] | N.O.[9] | N.I.[10] |
|---|---|---|---|---|---|---|---|---|---|---|
| **Bulbs, tubers, roots, vegetation or flowers, etc. and roots** | | | | | | | | | | |
| FE[11] | 0.929 (0.080) | -0.015 (-0.010) | 0.000 | 0.880 | 0.361 | ------- | ------- | 0.073 | 9 | ------- |
| RE[12] | 1.995 (1.160) | -0.112 (-0.740) | 0.550 | ------- | ------- | 0.000 | 0.010 | 0.073 | 9 | ------- |
| OLS | 1.995 (1.160) | -0.112 (-0.740) | 0.550 | ------- | ------- | ------- | ------- | 0.073 | 9 | ------- |
| DPD[13] | ------- | ------- | ------- | ------- | ------- | ------- | ------- | ------- | ------- | ------- |
| **Other live plants (including roots), cuttings and slips and mushroom spawn** | | | | | | | | | | |
| FE[11] | 14.922* (5.010) | -1.192* (-5.020) | 25.190* | 2.920* | -0.978 | ------- | ------- | 0.419 | 52 | ------- |
| RE[12] | 0.257 (0.360) | -0.021 (-0.370) | 0.140 | ------- | ------- | 1.410 | 25.770* | 0.419 | 52 | ------- |
| OLS | ------- | ------- | ------- | ------- | ------- | ------- | ------- | ------- | ------- | ------- |
| DPD[13] | 8.711* (3.200) | -0.683* (-3.220) | 22.470* | ------- | ------- | ------- | ------- | ------- | 22 | 5 |
| **Flowers and buds cut for branches and ornamental purposes, fresh, dried, etc.** | | | | | | | | | | |
| FE[11] | 11.467* (4.580) | -1.058* (-4.650) | 21.660* | 3.320* | -0.938 | ------- | ------- | 0.485 | 34 | ------- |
| RE[12] | 2.379 | -0.233 | 3.890* | ------- | ------- | 0.820 | 18.020* | 0.485 | 34 | ------- |

|  | | | | | | | | | |
|---|---|---|---|---|---|---|---|---|---|
|  | (1.770) | (-1.970) |  |  |  |  |  |  |  |  |
| OLS | ------- | ------- | ------- | ------- | ------- | ------- | ------- | ------- | ------- | ------- |
| DPD[13] | 15.450* | -1.427* | 41.910* | ------- | ------- | ------- | ------- | ------- | 15 | 5 |
|  | (4.630) | (-4.660) |  |  |  |  |  |  |  |  |
| **Foliage, branches and other parts of plants, without flowers and buttons, etc.** | | | | | | | | | | |
| FE[11] | 11.092* | -0.878* | 55.720* | 20.970* | -0.943 | ------- | ------- | 0.708 | 33 | ------- |
|  | (7.540) | (-7.460) |  |  |  |  |  |  |  |  |
| RE[12] | 6.002* | -0.473* | 30.920* | ------- | ------- | 4.040* | 24.360* | 0.708 | 33 | ------- |
|  | (5.590) | (-5.560) |  |  |  |  |  |  |  |  |
| OLS | ------- | ------- | ------- | ------- | ------- | ------- | ------- | ------- | ------- | ------- |
| DPD[13] | 18.512* | -1.450* | 38.690* | ------- | ------- | ------- | ------- | ------- | 16 | 5 |
|  | (6.000) | (-5.990) |  |  |  |  |  |  |  |  |

In the table 9 to stress the fact of the first estimations have results statistically less significant.

**Table 9. Results from the absolute convergence model for flowers import (percentage values relatively to the total of each country)**

|  | Const.[1] | Coef.[2] | F/Wald(mod.)[3] | F(Fe_OLS)[4] | Corr(u_i)[5] | F(Re_OLS)[6] | Hausman[7] | R[2][8] | N.O.[9] | N.I.[10] |
|---|---|---|---|---|---|---|---|---|---|---|
| **Bulbs, tubers, roots, vegetation or flowers, etc. and roots** | | | | | | | | | | |
| FE[11] | 0.632 | -0.533* | 8.230* | 1.380 | -0.791 | ------- | ------- | 0.286 | 34 | ------- |
|  | (1.720) | (-2.870) |  |  |  |  |  |  |  |  |
| RE[12] | -0.142 | -0.065 | 0.360 | ------- | ------- | 0.000 | 9.580* | 0.286 | 34 | ------- |
|  | (-0.510) | (-0.600) |  |  |  |  |  |  |  |  |
| OLS | -0.142 | -0.065 | 0.360 | ------- | ------- | ------- | ------- | 0.020 | 34 | ------- |
|  | (-0.510) | (-0.600) |  |  |  |  |  |  |  |  |
| DPD[13] | 1.569* | -1.150* | 50.420* | ------- | ------- | ------- | ------- | ------- | 15 | 5 |
|  | (4.360) | (-6.340) |  |  |  |  |  |  |  |  |
| **Other live plants (including roots), cuttings and slips and mushroom spawn** | | | | | | | | | | |
| FE[11] | 1.337* | -0.326* | 7.540* | 3.450* | -0.535 | ------- | ------- | 0.162 | 55 | ------- |
|  | (2.940) | (-2.750) |  |  |  |  |  |  |  |  |
| RE[12] | 0.630* | -0.149* | 4.030* | ------- | ------- | 1.760 | 3.610 | 0.162 | 55 | ------- |
|  | (2.130) | (-2.010) |  |  |  |  |  |  |  |  |
| OLS | 0.578* | -0.127* | 4.740* | ------- | ------- | ------- | ------- | 0.065 | 55 | ------- |
|  | (2.500) | (-2.180) |  |  |  |  |  |  |  |  |
| DPD[13] | 2.282* | -0.525* | 48.600* | ------- | ------- | ------- | ------- | ------- | 25 | 5 |
|  | (4.110) | (-3.680) |  |  |  |  |  |  |  |  |
| **Flowers and buds cut for branches and ornamental purposes, fresh, dried, etc.** | | | | | | | | | | |
| FE[11] | 2.940* | -1.138* | 35.350* | 2.890* | -0.805 | ------- | ------- | 0.549 | 43 | ------- |
|  | (5.410) | (-5.950) |  |  |  |  |  |  |  |  |
| RE[12] | 0.603 | -0.281* | 4.390* | ------- | ------- | 0.000 | 39.280* | 0.549 | 43 | ------- |
|  | (1.460) | (-2.100) |  |  |  |  |  |  |  |  |
| OLS | ------- | ------- | ------- | ------- | ------- | ------- | ------- | ------- | ------- | ------- |
| DPD[13] | 5.961* | -2.185* | 118.920* | ------- | ------- | ------- | ------- | ------- | 19 | 5 |
|  | (9.880) | (-10.250) |  |  |  |  |  |  |  |  |
| **Foliage, branches and other parts of plants, without flowers and buttons, etc.** | | | | | | | | | | |
| FE[11] | 1.241* | -0.992* | 35.880* | 3.690* | -0.772 | ------- | ------- | 0.521 | 47 | ------- |
|  | (4.660) | (-5.990) |  |  |  |  |  |  |  |  |
| RE[12] | 0.576 | -0.470* | 13.120* | ------- | ------- | 0.090 | 25.690* | 0.521 | 47 | ------- |
|  | (1.800) | (-3.620) |  |  |  |  |  |  |  |  |
| OLS | ------- | ------- | ------- | ------- | ------- | ------- | ------- | ------- | ------- | ------- |
| DPD[13] | 2.165* | -1.839* | 27.710* | ------- | ------- | ------- | ------- | ------- | 21 | 5 |
|  | (4.510) | (-5.010) |  |  |  |  |  |  |  |  |

For the table 10, of referring, again, the problem of the number of observations.

**Table 10. Results from the absolute convergence model for flowers export (percentage values relatively to the total of each country)**

|  | Const.[1] | Coef.[2] | F/Wald(mod.)[3] | F(Fe_OLS)[4] | Corr(u_i)[5] | F(Re_OLS)[6] | Hausman[7] | R[2][8] | N.O.[9] | N.I.[10] |
|---|---|---|---|---|---|---|---|---|---|---|
| **Bulbs, tubers, roots, vegetation or flowers, etc. and roots** | | | | | | | | | | |
| FE[11] | 0.339 | -0.003 | 0.000 | 1.530 | 0.273 | ------- | ------- | 0.049 | 10 | ------- |
|  | (0.400) | (-0.000) |  |  |  |  |  |  |  |  |
| RE[12] | 0.356 | -0.070 | 0.070 | ------- | ------- | 0.000 | 0.000 | 0.049 | 10 | ------- |
|  | (0.580) | (-0.260) |  |  |  |  |  |  |  |  |
| OLS | 0.417 | -0.145 | 0.410 | ------- | ------- | ------- | ------- | 0.071 | 10 | ------- |
|  | (0.890) | (-0.640) |  |  |  |  |  |  |  |  |
| DPD[13] | ------- | ------- | ------- | ------- | ------- | ------- | ------- | ------- | ------- | ------- |
| **Other live plants (including roots), cuttings and slips and mushroom spawn** | | | | | | | | | | |
| FE[11] | 3.752* | -0.901* | 44.580* | 2.810* | -0.851 | ------- | ------- | 0.567 | 50 | ------- |

| | | | | | | | | | |
|---|---|---|---|---|---|---|---|---|---|
| | (6.660) | (-6.680) | | | | | | | |
| RE[12] | 0.993* (2.940) | -0.239* (-2.960) | 8.780* | ------- | ------- | 0.000 | 37.420* | 0.567 | 50 | ------- |
| OLS | ------- | ------- | ------- | ------- | ------- | ------- | ------- | ------- | ------- | ------- |
| DPD[13] | 6.280* (5.800) | -1.523* (-5.860) | 40.180* | ------- | ------- | ------- | ------- | ------- | 23 | 5 |
| **Flowers and buds cut for branches and ornamental purposes, fresh, dried, etc.** | | | | | | | | | | |
| FE[11] | 2.491* (7.180) | -1.359* (-7.990) | 63.820* | 9.610* | -0.873 | ------- | ------- | 0.744 | 32 | ------- |
| RE[12] | 0.533 (1.280) | -0.357* (-2.180) | 4.750* | ------- | ------- | 0.060 | 482.300* | 0.744 | 32 | ------- |
| OLS | ------- | ------- | ------- | ------- | ------- | ------- | ------- | ------- | ------- | ------- |
| DPD[13] | 4.391* (3.460) | -2.296* (-3.540) | 24.580* | ------- | ------- | ------- | ------- | ------- | 14 | 5 |
| **Foliage, branches and other parts of plants, without flowers and buttons, etc.** | | | | | | | | | | |
| FE[11] | 3.309* (7.910) | -1.360* (-7.980) | 63.750* | 8.470* | -0.947 | ------- | ------- | 0.727 | 34 | ------- |
| RE[12] | 0.447 (1.540) | -0.171 (-1.700) | 2.880 | ------- | ------- | 1.050 | 75.260* | 0.727 | 34 | ------- |
| OLS | ------- | ------- | ------- | ------- | ------- | ------- | ------- | ------- | ------- | ------- |
| DPD[13] | 2.299 (1.420) | -0.964 (-1.460) | 4.210 | ------- | ------- | ------- | ------- | ------- | 16 | 5 |

Table 11 presents, again, the problem of the results statistically less significant for the first estimations.

**Table 11. Results from the absolute convergence model for flowers import (percentage values relatively to the total of each year)**

| | Const.[1] | Coef.[2] | F/Wald(mod.)[3] | F(Fe_OLS)[4] | Corr(u_i)[5] | F(Re_OLS)[6] | Hausman[7] | $R^2$[8] | N.O.[9] | N.I.[10] |
|---|---|---|---|---|---|---|---|---|---|---|
| **Bulbs, tubers, roots, vegetation or flowers, etc. and roots** | | | | | | | | | | |
| FE[11] | -0.446 (-1.840) | -0.751* (-3.480) | 12.120* | 1.420 | -0.879 | ------- | ------- | 0.327 | 36 | ------- |
| RE[12] | -0.298 (-1.190) | -0.123 (-1.220) | 1.490 | ------- | ------- | 0.000 | 10.850* | 0.327 | 36 | ------- |
| OLS | -0.298 (-1.190) | -0.123 (-1.220) | 1.490 | ------- | ------- | ------- | ------- | 0.014 | 36 | ------- |
| DPD[13] | 8.233 (1.500) | -0.735 (-1.600) | 6.560* | ------- | ------- | ------- | ------- | ------- | 10 | 5 |
| **Other live plants (including roots), cuttings and slips and mushroom spawn** | | | | | | | | | | |
| FE[11] | -1.087* (-5.460) | -1.207* (-6.610) | 43.710* | 4.050* | -0.962 | ------- | ------- | 0.529 | 55 | ------- |
| RE[12] | -0.177 (-0.790) | -0.147* (-2.120) | 4.510* | ------- | ------- | 1.160 | 39.350* | 0.529 | 55 | ------- |
| OLS | ------- | ------- | ------- | ------- | ------- | ------- | ------- | ------- | ------- | ------- |
| DPD[13] | -1.682* (-6.340) | -2.243* (-6.530) | 76.030* | ------- | ------- | ------- | ------- | ------- | 25 | 5 |
| **Flowers and buds cut for branches and ornamental purposes, fresh, dried, etc.** | | | | | | | | | | |
| FE[11] | -0.443* (-2.450) | -0.854* (-5.040) | 25.410* | 2.970* | -0.910 | ------- | ------- | 0.451 | 45 | ------- |
| RE[12] | -0.074 (-0.310) | -0.221* (-2.700) | 7.290* | ------- | ------- | 0.160 | 18.190* | 0.451 | 45 | ------- |
| OLS | ------- | ------- | ------- | ------- | ------- | ------- | ------- | ------- | ------- | ------- |
| DPD[13] | -0.718* (-2.890) | -1.462* (-4.020) | 40.340* | ------- | ------- | ------- | ------- | ------- | 19 | 5 |
| **Foliage, branches and other parts of plants, without flowers and buttons, etc.** | | | | | | | | | | |
| FE[11] | -0.117 (-0.990) | -0.683* (-4.730) | 22.370* | 2.930* | -0.890 | ------- | ------- | 0.411 | 46 | ------- |
| RE[12] | -0.130 (-0.750) | -0.139 (-1.950) | 3.810* | ------- | ------- | 0.000 | 18.740* | 0.411 | 46 | ------- |
| OLS | ------- | ------- | ------- | ------- | ------- | ------- | ------- | ------- | ------- | ------- |
| DPD[13] | -0.104 (-0.950) | -1.315* (-6.600) | 52.290* | ------- | ------- | ------- | ------- | ------- | 20 | 5 |

In the table 12, more one time the problem of the number of observations in the first estimations.

**Table 12. Results from the absolute convergence model for flowers export (percentage values relatively to the total of each year)**

|  | Const.[1] | Coef.[2] | F/Wald(mod.)[3] | F(Fe_OLS)[4] | Corr(u_i)[5] | F(Re_OLS)[6] | Hausman[7] | $R^2$[8] | N.O.[9] | N.I.[10] |
|---|---|---|---|---|---|---|---|---|---|---|
| **Bulbs, tubers, roots, vegetation or flowers, etc. and roots** | | | | | | | | | | |
| **FE**[11] | 1.590 (1.650) | -0.601 (-0.900) | 0.810 | 0.760 | -0.897 | ------- | ------- | 0.455 | 9 | ------- |
| **RE**[12] | 1.084 (2.490*) | -0.215 (-1.770) | 3.130 | ------- | ------- | 0.440 | 0.350 | 0.455 | 9 | ------- |
| **OLS** | 1.090 (2.570*) | -0.217 (-1.820) | 3.320 | ------- | ------- | ------- | ------- | 0.225 | 9 | ------- |
| **DPD**[13] | ------- | ------- | ------- | ------- | ------- | ------- | ------- | ------- | ------- | ------- |
| **Other live plants (including roots), cuttings and slips and mushroom spawn** | | | | | | | | | | |
| **FE**[11] | -0.044 (-0.430) | -1.167* (-4.910) | 24.120* | 2.850* | -0.978 | ------- | ------- | 0.408 | 52 | ------- |
| **RE**[12] | -0.061 (-0.400) | -0.020 (-0.370) | 0.140 | ------- | ------- | 1.330 | 24.640* | 0.408 | 52 | ------- |
| **OLS** | ------- | ------- | ------- | ------- | ------- | ------- | ------- | ------- | ------- | ------- |
| **DPD**[13] | 0.127 (0.088) | -0.705* (-3.210) | 22.030* | ------- | ------- | ------- | ------- | ------- | 22 | 5 |
| **Flowers and buds cut for branches and ornamental purposes, fresh, dried, etc.** | | | | | | | | | | |
| **FE**[11] | -0.150 (-0.870) | -0.981* (-4.110) | 16.900* | 2.870* | -0.932 | ------- | ------- | 0.424 | 34 | ------- |
| **RE**[12] | -0.120 (-0.380) | -0.228 (-1.920) | 3.670 | ------- | ------- | 0.620 | 13.220* | 0.424 | 34 | ------- |
| **OLS** | ------- | ------- | ------- | ------- | ------- | ------- | ------- | ------- | ------- | ------- |
| **DPD**[13] | -0.281 (-1.610) | -1.462* (-4.580) | 32.900* | ------- | ------- | ------- | ------- | ------- | 15 | 5 |
| **Foliage, branches and other parts of plants, without flowers and buttons, etc.** | | | | | | | | | | |
| **FE**[11] | 0.461* (4.880) | -0.953* (-6.790) | 46.160* | 12.790* | -0.955 | ------- | ------- | 0.667 | 33 | ------- |
| **RE**[12] | 0.205 (0.580) | -0.404* (-4.630) | 21.480* | ------- | ------- | 1.860 | 24.630* | 0.667 | 33 | ------- |
| **OLS** | ------- | ------- | ------- | ------- | ------- | ------- | ------- | ------- | ------- | ------- |
| **DPD**[13] | 1.121* (6.130) | -1.556* (-6.500) | 48.480* | ------- | ------- | ------- | ------- | ------- | 16 | 5 |

Tables 13, 14 and 15, present the same data, but more aggregated. Give us another perspective of the data.

**Table 13. Results from the absolute convergence model for all flowers (absolute values)**

|  | Const.[1] | Coef.[2] | F/Wald(mod.)[3] | F(Fe_OLS)[4] | Corr(u_i)[5] | F(Re_OLS)[6] | Hausman[7] | $R^2$[8] | N.O.[9] | N.I.[10] |
|---|---|---|---|---|---|---|---|---|---|---|
| **Import** | | | | | | | | | | |
| **FE**[11] | 9.590* (9.000) | -0.833* (-9.110) | 82.930* | 2.390* | -0.927 | ------- | ------- | 0.390 | 182 | ------- |
| **RE**[12] | 1.138* (2.830) | -0.106* (-3.140) | 9.880* | ------- | ------- | 2.220 | 73.110* | 0.390 | 182 | ------- |
| **OLS** | ------- | ------- | ------- | ------- | ------- | ------- | ------- | ------- | ------- | ------- |
| **DPD**[13] | 19.238* (9.590) | -1.653* (-9.700) | 147.870* | ------- | ------- | ------- | ------- | ------- | 81 | 5 |
| **Export** | | | | | | | | | | |
| **FE**[11] | 12.048* (7.820) | -1.001* (-7.800) | 60.890* | 3.580* | -0.956 | ------- | ------- | 0.409 | 128 | ------- |
| **RE**[12] | 2.238* (3.720) | -0.185* (-3.720) | 13.830* | ------- | ------- | 0.130 | 47.580* | 0.409 | 128 | ------- |
| **OLS** | ------- | ------- | ------- | ------- | ------- | ------- | ------- | ------- | ------- | ------- |
| **DPD**[13] | 14.249* (6.390) | -1.160* (-6.390) | 91.770* | ------- | ------- | ------- | ------- | ------- | 56 | 5 |

**Table 14. Results from the absolute convergence model for all flowers (percentage values relatively to the total of each country)**

|  | Const.[1] | Coef.[2] | F/Wald(mod.)[3] | F(Fe_OLS)[4] | Corr(u_i)[5] | F(Re_OLS)[6] | Hausman[7] | $R^2$[8] | N.O.[9] | N.I.[10] |
|---|---|---|---|---|---|---|---|---|---|---|
| **Import** | | | | | | | | | | |
| **FE**[11] | 1.940* (8.550) | -0.810* (-9.340) | 87.300* | 2.570* | -0.854 | ------- | ------- | 0.407 | 179 | ------- |
| **RE**[12] | 0.211 | -0.115* | 6.220* | ------- | ------- | 0.670 | 89.410* | 0.407 | 179 | ------- |

|  |  | (1.470) | (-2.490) |  |  |  |  |  |  |  |  |
|---|---|---|---|---|---|---|---|---|---|---|---|
| OLS |  | ------- | ------- | ------- | ------- | ------- | ------- | ------- | ------- | ------- | ------- |
| DPD[13] |  | 3.557* (11.000) | -1.440* (-11.480) | 161.100* | ------- | ------- | ------- | ------- | ------- | 80 | 5 |
| **Export** |  |  |  |  |  |  |  |  |  |  |  |
| FE[11] |  | 3.586* (11.970) | -1.263* (-12.110) | 146.580* | 5.750* | -0.946 | ------- | ------- | 0.628 | 126 | ------- |
| RE[12] |  | 0.610* (3.020) | -0.218* (-3.650) | 13.340* | ------- | ------- | 1.340 | 149.030* | 0.628 | 126 | ------- |
| OLS |  | ------- | ------- | ------- | ------- | ------- | ------- | ------- | ------- | ------- | ------- |
| DPD[13] |  | 5.338* (6.590) | -1.845* (-6.610) | 83.070* | ------- | ------- | ------- | ------- | ------- | 55 | 5 |

**Table 15. Results from the absolute convergence model for all flowers (percentage values relatively to the total of each year)**

|  | Const.[1] | Coef.[2] | F/Wald(mod.)[3] | F(Fe_OLS)[4] | Corr(u_i)[5] | F(Re_OLS)[6] | Hausman[7] | $R^2$ [8] | N.O.[9] | N.I.[10] |
|---|---|---|---|---|---|---|---|---|---|---|
| **Import** |  |  |  |  |  |  |  |  |  |  |
| FE[11] | -0.466* (-5.380) | -0.848* (-9.480) | 89.840* | 2.430* | -0.919 | ------- | ------- | 0.409 | 182 | ------- |
| RE[12] | -0.132 (-1.400) | -0.123* (-3.570) | 12.770* | ------- | ------- | 2.010 | 77.080* | 0.409 | 182 | ------- |
| OLS | ------- | ------- | ------- | ------- | ------- | ------- | ------- | ------- | ------- | ------- |
| DPD[13] | -0.739* (-7.790) | -1.714* (-9.820) | 161.170* | ------- | ------- | ------- | ------- | ------- | 81 | 5 |
| **Export** |  |  |  |  |  |  |  |  |  |  |
| FE[11] | 0.211* (2.910) | -0.986* (-7.960) | 63.290* | 3.590* | -0.953 | ------- | ------- | 0.418 | 128 | ------- |
| RE[12] | 0.042 (0.270) | -0.196* (-3.900) | 15.180* | ------- | ------- | 0.320 | 48.570* | 0.418 | 128 | ------- |
| OLS | ------- | ------- | ------- | ------- | ------- | ------- | ------- | ------- | ------- | ------- |
| DPD[13] | 0.566* (5.810) | -1.287* (-7.570) | 99.180* | ------- | ------- | ------- | ------- | ------- | 56 | 5 |

Using tests of stationary, it is analyzed the volatility of the data, and it is verified that all the results for all the tests are not significant, so we can conclude about no stationary of the data.

## 4. Conclusions

The most important relationship of Portugal with the world in the international trade of flowers is with the Europe, namely with Netherlands and Spain. So, there is a tendency to trade flowers with the neighbors, to economize costs of transport. Portugal import and export in particular other live plants (including roots), cuttings and slips and mushroom spawn.

There are strong signs of convergence in this international trade of flowers. In the majority of the results the values have statistical significance. So, we have convergence in the import and export of flowers.

In terms of stationary of the results we find a preoccupant volatility of the values for the different forms of flowers imports and exports. These results can compromise the conclusions presented before. This means that we do not have a coherent Policy for the international trade of flowers.

When we speak about tradable products like the flowers, is important to have a good policy for the international trade, because, as said by the Keynesian theory the external demand (exports) are the engine of the national product.

Anyway, this is a contribution to the international trade flowers sector understanding, using Portugal as the central country.